\documentclass[
    ,final            % use final for the camera ready runs
  ,draft            % use draft while you are working on the paper
  ,numberedheadings % uncomment this option for numbered sections
  ]
  {aipproc}

\usepackage[mathscr]{eucal}
 \usepackage{amsmath, amssymb}

\def\dmf{\dot{\mathfrak{M}}}

\newcommand{\be}{\begin{equation}}
\newcommand{\ee}{\end{equation}}
\newcommand{\bdm}{\begin{displaymath}}
\newcommand{\edm}{\end{displaymath}}
\newcommand{\vect}[1]{\mathbf{#1}}
\layoutstyle{6x9}

\begin{document}

%\vspace{-5.5cm}
\noindent Published in {\it Astronomy Reports}, Volume\,58, Issue\,6, pp.\,376--385
(2014)

\vspace{1cm}

\title{Spin evolution of long-period X-ray pulsars}

\classification{97.10.Gz, 97.80.Jp, 95.30.Qd}
\keywords{Accretion and accretion disks, X-ray binaries, neutron star, pulsars, magnetic field}

\author{N.R.\,Ikhsanov\footnote{Saint-Petersburg State University,
Universitetsky pr., 28, Saint Petersburg 198504, Russia}}{
  address={Pulkovo Observatory, Pulkovskoe Shosse 65, Saint-Petersburg 196140, Russia}
}

\author{Y.S.\,Likh}{
  address={Pulkovo Observatory, Pulkovskoe Shosse 65, Saint-Petersburg 196140, Russia}
}

\author{N.G.\,Beskrovnaya}{
  address={Pulkovo Observatory, Pulkovskoe Shosse 65, Saint-Petersburg 196140, Russia}
}

\begin{abstract}
Spin evolution of X-ray pulsars in High Mass X-ray Binaries (HMXBs) is discussed
under various assumptions about the geometry and physical parameters of the accretion
flow. The torque applied to the neutron star from the accretion flow and equilibrium
period of the pulsars are evaluated. We show that the observed spin evolution of the
pulsars can be explained in terms of a scenario in which the neutron star accretes
material from a magnetized stellar wind.
\end{abstract}

\maketitle

%%%%%%%%%%%%%%%%%%%%%%%%%%%%%%%%%%%%%%%%%%%%
%% MAINMATTER
%%%%%%%%%%%%%%%%%%%%%%%%%%%%%%%%%%%%%%%%%%%%

   \section{Introduction}

Long-period X-ray pulsars  constitute a subclass of X-ray sources displaying regular
pulsations with periods in the range from a few tens to a few thousand seconds in their
X-ray emission. This subclass now includes more than 40 objects
\cite{Liu-etal-2006, Raguzova-Popov-2005}. The majority of them are identified
with X-ray binary systems containing an early-type star and a strongly magnetized neutron star.
The X-ray emission of these systems is accretion-driven, i.e. is produced due to accretion of matter
onto the neutron star surface in the region of its magnetic poles. Regular intensity variations
of radiation emitted from these regions occur with a period equal to the spin period of the neutron star
whose magnetic axis is inclined to its rotational axis  (see \cite{Lipunov-1987}
and references therein).

\begin{figure}
\includegraphics[width=12cm]{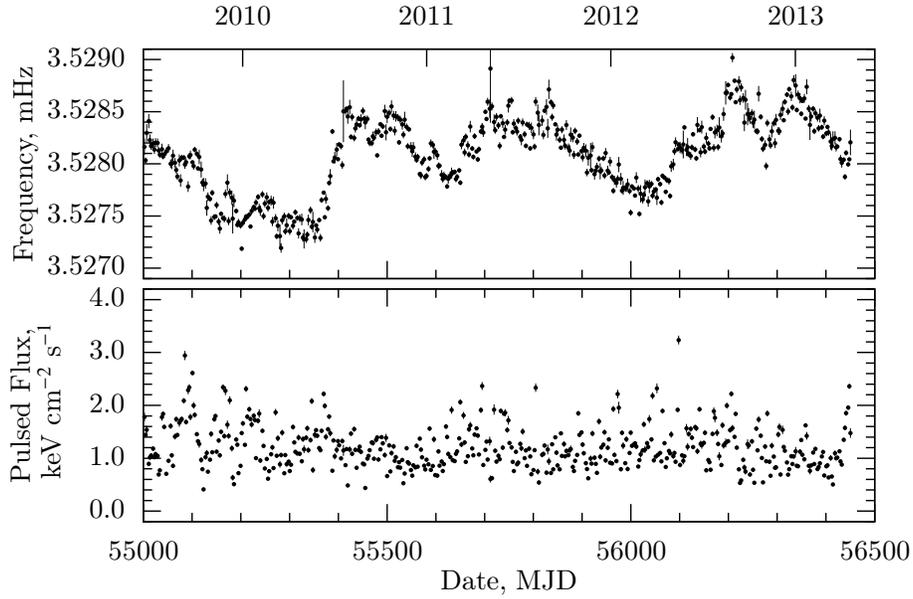}
\caption{Spin evolution of the long-period X-ray pulsar
Vela~X--1(top) and its X-ray flux variations in the energy range  $12-50$\,keV (bottom)
observed  with the X-ray telescope onboard Fermi GMB \cite{GBM-VELA}.}
 \label{fig-1}
 \end{figure}

The periods of long-period X-ray pulsars are not constant. They undergo, as a rule, chaotic variations
over timescales from several days to several years while  the average period does not change significantly
 (see Fig.\,\ref{fig-1}). Such a behavior is  explained by the  exchange of angular momentum between a star and surrounding
 material which is governed by the equation
 \cite{Lipunov-1987}
 \be\label{main}
 2 \pi I \dot{\nu} = K_{\rm su} - K_{\rm sd}.
 \ee
Here  $I$ is the moment of inertia of a neutron star, and $\nu = 1/P_{\rm s}$
is the frequency of its axial rotation with the period
$P_{\rm s}$. The right side of the equation is the difference between the spin-up,
 $K_{\rm su}$,  and spin-down,  $K_{\rm sd}$, torques applied to the star from the accretion flow.
 Evaluation of these torques is the key problem in the modeling of the pulsar spin evolution.
 The spin period of a neutron star at which a balance between spin-up and spin-down torques
 is achieved  ($K_{\rm su} = K_{\rm sd}$), is called the equilibrium spin period,  $P_{\rm eq}$.

Until recently the spin evolution of long-period X-ray pulsars has been  modeled
exclusively under the assumption that the mass transfer beyond the magnetospheric
boundary of a neutron star proceeds through either a Keplerian disk  or  a
quasi-spherical flow. In particular,  this approach has led to the conclusion that the
absolute value of the spin-down torque is limited  by inequality
 $|K_{\rm sd}| \leq
|K_{\rm sd}^{(0)}|$, where \cite{Shakura-1975, Menou-etal-1999, Shakura-etal-2012}
 \be\label{ksd0}
 |K_{\rm sd}^{(0)}| = k_{\rm t}\,\dmf\,\omega_{\rm s}\,r_{\rm A}^2.
 \ee
The accretion rate onto the surface of a  neutron star of the radius
 $R_{\rm ns}$ and mass $M_{\rm
ns}$, which appears in this equation, $\dmf = L_{\rm X} R_{\rm ns}/GM_{\rm ns}$, is evaluated from the X-ray luminosity of the pulsar, $L_{\rm X}$.
Here  $\omega_{\rm s} = 2 \pi \nu$ is the angular velocity of the neutron star and
 $k_{\rm t}$ is  a dimensionless parameter of an order of unity. The magnetospheric radius of a neutron star within this approach coincides
with the canonical Alfven radius,
  \be
 r_{\rm A} = \left(\frac{\mu^2}{\dmf (2 GM_{\rm ns})^{1/2}}\right)^{2/7},
 \ee
at which the magnetic pressure due to dipole magnetic field of the neutron star, $\mu^2/(2
\pi r^6)$, reaches the ram pressure of free-falling gas,
 $E_{\rm ram}(r) = \rho(r) v_{\rm
ff}^2(r)$. The density and velocity of the gas in the spherical accretion flow
at a given radius are determined as
$\rho(r) = \dmf/[4 \pi r^2 v_{\rm ff}(r)]$ and $v_{\rm ff}(r) =
\left(2GM_{\rm ns}/r\right)^{1/2}$, respectively, and the dipole magnetic moment of a neutron star,
 $\mu = (1/2) B_* R_{\rm ns}^3$, can be written in terms of the surface magnetic field,  $B_*$.

Studies of  the most fully investigated long-period X-ray pulsars (see Table\,\ref{tab-1})
have shown, however, that their observed spin-down rates, $|\dot{\nu}_{\rm sd}^{\rm obs}|$, occasionally (see Table~\ref{tab-2})
significantly exceed the maximum possible value,   $|\dot{\nu}_{\rm sd}^{(0)}| \leq
|K_{\rm sd}^{(0)}|/(2 \pi I)$, predicted within the canonical models (see Table\,\ref{tab-3}).
This discrepancy between  theoretical and empirical data has been stated many times in the literature
starting from 1975 (see, e.g.,
 \cite{Shakura-1975, Lipunov-Shakura-1976,
Lipunov-1982a, Boerner-etal-1987, Finger-etal-2010}).
It can be avoided in the frame of traditional accretion scenarios only under assumption
that the surface magnetic field of these neutron stars is
 $10^{14} - 10^{15}$\,G which exceeds by two orders of magnitude the value inferred from observations of cyclotron
 line in their X-ray spectra.

Recently, however, doubts have been pronounced on a feasibility of an assumption about a presence of super-strong magnetic fields
on the surface of long-period X-ray pulsars  \cite{Ikhsanov-Finger-2012, Ikhsanov-2012}. Investigation of the long-period X-ray pulsar GX\,301-2 has revealed that a controversy between theoretical predictions and observational data
arises only when one attempts to explain the episodes of its rapid breaking. The magnetic field of the neutron star estimated from observations of its rapid spin-up episodes agrees with the value obtained through observations of the cyclotron feature in the X-ray spectra of this source (see \cite{Ikhsanov-Finger-2012} and references therein). An attempt to describe the episode of rapid spin-down of the neutron star SPX\,1062 within the conventional model leads to a more paradoxical result: The Alfven radius of the neutron star turns out to exceed its corotation radius, $r_{\rm cor} = \left(GM_{\rm ns}/\omega_{\rm s}^2\right)^{1/3}$, that excludes a possibility of stationary matter accretion onto its surface and  obscures the nature of this X-ray source \cite{Ikhsanov-2012}.

These results indicate that the spin-down torque exerted on a neutron star by the accretion flow is likely to exceed $|K_{\rm sd}^{(0)}|$, and underestimation  of this parameter may be connected with undue simplification of commonly used accretion
scenarios. Most doubt is cast upon a validity of an assumption that the magnetic field in the material captured by a neutron star from the wind of its massive companion is negligibly weak. At early stages of the accretion theory development this notion was grounded on a common belief about a relatively weak magnetization of the massive early-type stars. In contrast to solar-type stars and late-type dwarfs, these objects do not contain sub-photospheric convective zones in which the magnetic field can be amplified in the dynamo process. Spectropolarimetric observations of the last decade have shown, however, that the large-scale magnetic fields of  O/B stars can reach hundreds and even a thousand gauss \cite{Hubrig-etal-2006, Oksala-etal-2010, Martins-etal-2010}.
Magnetic pressure in the stellar wind of such a star remains comparable to the thermal pressure of the outflowing plasma even at large distances from the star \cite{Ikhsanov-Beskrovnaya-2012, Ikhsanov-Beskrovnaya-2013} and can essentially affect the pattern of matter flow inside Bondi radius, $r_{\rm G} = 2 GM_{\rm ns}/v_{\rm rel}^2$, of its degenerate companion moving  with velocity $v_{\rm rel}$ relative to the stellar wind.

%%%%%%%%%%%%%%%%%% Таблица 1 %%%%%%%%%%%%%%%%%%%%%

\begin{table}
\bigskip
\begin{tabular}{lccccccc}
 \noalign{\smallskip}
  \hline
\noalign{\smallskip}
 Name &
 $P_{\rm s}\ ^*$,\,s &
 $P_{\rm orb}$,\,d &
 $L_{\rm X}\ ^*$,\,$10^{36}$\,erg\,s$^{-1}$ &
 $B\,^{**}$,\,$10^{12}$\,G &
 Sp.type &
 d,\,kpc &
 Ref. \\
 \noalign{\smallskip}
\hline
\noalign{\smallskip}
 OAO~1657--415 &
 37   &
 10.4 &
 3    &
 3.2  &
 Ofpe &
 6.4  &
 \cite{Chakrabarty-etal-2002, Karino-2007} \\
 Vela~X--1 &
 283  &
 9    &
 4    &
 2.6  &
 B0.5~Ib &
 2    &
 \cite{Nagase-etal-1986, Kreykenbohm-etal-2002} \\
 4U~1907+09 &
 441   &
 8     &
 2     &
 2.1   &
 O8-9~Ia &
 4     &
 \cite{Cox-etal-2005, Cusumano-1998} \\
 4U~1538--52 &
 525   &
 4     &
 2     &
 2.3   &
 B0~Iab &
 4.5   &
 \cite{Parkes-etal-1978, Coburn-etal-2002} \\
 GX~301--2 &
 685   &
 41.5  &
 10    &
 4     &
 B1~Ia& 3  &
 \cite{Chichkov-etal-1995, Kreykenbohm-2004} \\
 X~Persei &
 837   &
 250   &
 0.1   &
 3.3   &
 B0~Ve& 1 &
 \cite{Palombara-2007, Coburn-etal-2001} \\
 \noalign{\smallskip}
  \hline
   \noalign{\smallskip}
   \multicolumn{8}{l}{$^*$~~The value averaged over all published data} \\
   \multicolumn{8}{l}{$^{**}$~This value  is estimated through observations} \\
     \multicolumn{8}{l}{of the cyclotron line in the X-ray spectrum of an object}
 \end{tabular}
 \caption{Long-period X-ray pulsars}
 \bigskip
 \label{tab-1}
 \end{table}

%%%%%%%%%%%%%%%%%%%%%%%%%%%%%%%%%%%%%%%%%%%%%%%%%%%%%%%%%%%%%%%%%%

In this paper we estimate the spin-down torque applied to a neutron star from the
accretion flow in the frame of a  model problem on a sphere rotation in  viscous medium
\cite{Lipunov-1987}. We show that for the parameters of interest the torque is
increasing with the sphere's radius decreasing as $K_{\rm sd} \propto r^{-3/2}$. Its
estimate (see Eq.~\ref{ksd2}) corresponds to previously obtained solution, $K_{\rm
sd}^{(0)}$, for  $r_{\rm m} = r_{\rm A}$ and significantly exceeds this value if the
radius at which the flow enters the stellar magnetic field is less than the Alfven
radius, $r_{\rm m} < r_{\rm A}$. In particular, the observed spin-down rate of the
long-period X-ray pulsars within our approach can be obtained if
 $r_{\rm m} \sim (0.1-0.5)\,r_{\rm A}$ (see
Table\,\ref{tab-4}). This conclusion makes it difficult to model the accretion picture
in terms of Keplerian disk or quasi-spherical flow (see Section\,\ref{spin-down}) and
force us to consider alternative scenarios. One of them is the model of accretion from
the magnetized wind, first suggested by Shvartsman \cite{Shvartsman-1971} and further
elaborated by Bisnovatyi-Kogan and Ruzmaikin \cite{Bisnovatyi-Kogan-Ruzmaikin-1974,
Bisnovatyi-Kogan-Ruzmaikin-1976}. Within this approach, a neutron star accretes material
from a non-Keplerian magnetic slab surrounding its magnetosphere. The radial motion of
the matter inside the slab occurs in the diffusion regime on the annihilation time scale
of the intrinsic magnetic field of the accretion flow. We show that the magnetospheric
radius of a neutron star under these conditions can be significantly less than the
canonical Alfven radius and close to the value inferred from observed spin-down rates of
long-period X-ray pulsars (Section\,\ref{slab}). Equilibrium period of a neutron star
expected in different accretion scenarios (Keplerian disk, quasi-spherical flow and
magnetic slab) is estimated in Section\,\ref{peq}. We briefly discuss basic conclusions
of the paper in Section~\ref{conclusions}.

%%%%%%%%%%%%%%%%%%   Таблица 2 %%%%%%%%%%%%%%%%%%%%%

\begin{table}
\begin{tabular}{lccccc}
 \noalign{\smallskip}
  \hline
\noalign{\smallskip}
 Name &
 \hspace{5mm} $t_0$,\,MJD &
 \hspace{5mm} $t_1$,\,MJD &
 \hspace{5mm} $\bigtriangleup$t,\,d &
 \hspace{5mm} $|\dot{\nu}_{\rm sd}^{\rm obs}|$,\,Hz\,s$^{-1}$ &
 \hspace{5mm} Ref. \\
 \noalign{\smallskip}
 \hline
 \noalign{\smallskip}
OAO~1657--415 & 50680 & 50687 &  7 & $3\times10^{-12}$ & \cite{Baykal-2000} \\
Vela~X--1     & 44306 & 44320 &   14& $3\times10^{-13}$ & \cite{Deeter-etal-1987} \\
4U~1907+09    & 54280 & 55600 & 1320& $4\times10^{-14}$ & \cite{Sahiner-etal-2012} \\
4U~1538--52   & 45514 & 45522 &    8& $2\times10^{-13}$ & \cite{Makishima-etal-1987}\\
GX~301--2     & 54300 & 54710 &  410 & $10^{-13}$ & \cite{Evangelista-etal-2010}\\
X~Persei      & 43413 & 43532 &  118& $2\times10^{-14}$ & \cite{Weisskopf-etal-1984} \\
 \noalign{\smallskip}
 \hline
\end{tabular}
 \caption{Observed spin-down rates}
 \bigskip
 \label{tab-2}
\end{table}

%%%%%%%%%%%%%%%%%%%%%%%%%%%%%%%%%%%%%%%%%%%%%%%%%%%%%%%%%%%

\section{Spin-down torque}\label{spin-down}

Following Lipunov \cite{Lipunov-1987}, the spin-down torque exerted on a neutron star by
the accretion flow can be estimated in the frame of model problem in which a sphere of
radius $r_{\rm m}$ is rotating with angular velocity $\omega_{\rm s}$ in a medium with
viscosity $\nu_{\rm t}$. The absolute value of spin-down torque in this case is
evaluated as
 \be\label{ksd1}
 |K_{\rm sd}| = \nu_{\rm t}\,S_{\rm eff}\,\rho_0\,v_{\phi},
 \ee
where $S_{\rm eff}$ is the effective area of interaction between the sphere and
accretion flow of  density  $\rho_0 = \rho(r_{\rm m})$. A parameter
 \be\label{vphi}
v_{\phi} = \omega_{\rm s} r_{\rm m} \left(1 -
\frac{\Omega_0}{\omega_{\rm s}}\right),
 \ee
coming into this expression defines the azimuthal velocity component of the
magnetospheric boundary relative to the surrounding gas rotating with angular velocity
$\Omega_0 = \Omega(r_{\rm m})$ (assuming for simplicity ${\Omega} \parallel \vect{\omega}_{\rm s}$).
Combining these expressions we get the spin-down torque in a form
  \be
 \vect{K_{\rm sd}} = \vect{r_{\rm m}} \times \vect{F},
 \ee
corresponding to the definition of this parameter as the cross product of force exerted
on a sphere by surrounding gas,
 \be
|\vect{F}| = \nu_{\rm t} S_{\rm eff} \rho_0 \omega_{\rm s} \left(1 -
\frac{\Omega_0}{\omega_{\rm s}}\right),
 \ee
and the lever arm, $\vect{r_{\rm m}}$.

The effective area of interaction within approximation of azimuthal symmetry of the
accretion flow can be expressed as
 \be\label{seff}
 S_{\rm eff} = 4 \pi\,r_{\rm m}\,h_{\rm z}(r_{\rm m}),
 \ee
where
 \be
 h_{\rm z}(r_{\rm m}) = \frac{r_{\rm m}^2\ c_{\rm s}^2(r_{\rm m})}{GM_{\rm ns}}
 \ee
is the height of homogeneous atmosphere which determines the scale of accretion flow in
the direction, perpendicular to the plane of its rotation.  Here $c_{\rm s}(r_{\rm m})$
is the sound speed in the surrounding gas. The case of $c_{\rm s} = v_{\rm k}(r_{\rm
m})$, where $v_{\rm k}(r_{\rm m}) = \left(GM_{\rm ns}/r_{\rm m}\right)^{1/2}$ is the
Keplerian velocity at the magnetospheric boundary, corresponds to the spherical geometry
of the accretion flow in which the effective area of interaction reaches the maximum
value $4 \pi r_{\rm m}^2$.

Plasma density at the magnetospheric boundary in general case can be estimated from the
balance of magnetic pressure due to the neutron star dipole field,
 $\mu^2/(2 \pi r_{\rm m}^6)$, and the gas pressure, $\rho_0 c_{\rm s}^2(r_{\rm m})$. This leads us to expression
 \be\label{rho0}
 \rho_0 = \frac{\mu^2}{2 \pi r_{\rm m}^6 c_{\rm s}^2(r_{\rm m})}.
 \ee
Putting Eqs.~(\ref{seff}--\ref{rho0}) into (\ref{ksd1}), we find
 \be
 |K_{\rm sd}| = \nu_{\rm t} \left(\frac{\mu^2}{r_{\rm m}^2}
 \frac{\omega_{\rm s}}{GM_{\rm ns}}\right) \left(1 - \frac{\Omega_0}{\omega_{\rm s}}\right).
 \ee
Finally, adopting the turbulent viscosity of the medium,
 $\nu_{\rm t} = k_{\rm
t} \ell_{\rm t} v_{\rm t}$, and taking into account that the scale and velocity of the
turbulent motions at the magnetospheric boundary under conditions of interest are
limited by the inequalities  $\ell_{\rm t} \leq r_{\rm m}$ and $v_{\rm t} \leq v_{\rm
k}(r_{\rm m})$, we derive the expression
 \be\label{ksd2}
 |K_{\rm sd}| = k_{\rm t}\ \frac{\mu^2}{\left(r_{\rm m}
 r_{\rm cor}\right)^{3/2}} \left(1 - \frac{\Omega_0}{\omega_{\rm s}}\right),
 \ee
where $0 < k_{\rm t} <1$ is the dimensionless parameter.

The Eq.~(\ref{ksd2}) is a generalized prescription of the spin-down torque
applied to the neutron star from the accretion flow. The previous estimate of this
parameter,  $|K_{\rm sd}^{(0)}|$, is a particular case of this expression which can be
obtained substituting  $r_{\rm m} = r_{\rm A}$ provided the angular velocity of the
accretion flow at the boundary is zero,  $\Omega_0 = 0$. If the turbulent velocity
 in the accretion flow at the magnetospheric boundary does not exceed  $\omega_{\rm s}
r_{\rm m}$, we come to the expression
 \be\label{ksdt}
 |K_{\rm sd}^{\rm (t)}| = |K_{\rm sd}^{(0)}|
 \left(\frac{\omega_{\rm s} r_{\rm m}}{v_{\rm k}(r_{\rm m})}\right) =
 k_{\rm t} \frac{\mu^2}{r_{\rm cor}^3},
 \ee
presented earlier in \cite{Lynden-Bell-Pringle-1974, Lipunov-1982}.

Thus,  our evaluation of the spin-down torque agrees with previous estimates of this
parameter obtained under assumption that the neutron star accretes material from a
Keplerian disk or  turbulent quasi-spherical flow. However, it indicates that
the magnitude of the spin-down torque can be significantly larger in the frame of
accretion scenario in which the magnetospheric radius of the neutron star is smaller than
the canonical Alfven radius. In particular, the observed spin-down rates of the
long-period X-ray pulsars (see Table~\ref{tab-2}) can be explained using
Eq.~(\ref{ksd2}) provided $r_{\rm m} \leq r_0$, where
 \be\label{r0}
 r_0 = \frac{1}{r_{\rm cor}} \left(\frac{k_{\rm t} \mu^2}
 {2 \pi I \dot{\nu}_{\rm sd}}\right)^{2/3}.
 \ee
The estimates of $r_0$, calculated  for the parameters of the most
fully studied long-period X-ray pulsars and  $k_{\rm t} = 1$  are significantly smaller
than the Alfven radii of these objects (see Table~\ref{tab-4}) and, hence, are less than the magnetospheric
radii of neutron stars expected in case of accretion from a Keplerian disk or
quasi-spherical flow. This circumstance force us to address alternative accretion
models, one of which considers accretion from a magnetized stellar wind.

 \section{Accretion from a non-Keplerian magnetic slab}\label{slab}

As was first shown by Shvartsman \cite{Shvartsman-1971}, the distance at which the
spherical accretion flow with magnetic field  $B_{\rm f}$ can approach the compact star
in the free-fall regime is defined by the expression
    \be\label{rsh}
 R_{\rm sh} = \beta_0^{-2/3} \left(\frac{c_{\rm s}(r_{\rm G})}
 {v_{\rm rel}}\right)^{4/3} r_{\rm G}.
 \ee
Here $\beta_0 = E_{\rm th}(r_{\rm G})/E_{\rm m}(r_{\rm G})$ is the ratio of the thermal
pressure, $E_{\rm th} = \rho c_{\rm s}^2$, to the magnetic pressure, $E_{\rm m} = B_{\rm
f}^2/8 \pi$, in the material captured by a compact star from the wind of its companion
at the Bondi radius, and  $c_{\rm s}$ is the sound speed in the accretion flow. At the
radius  $R_{\rm sh}$, a so called Shvartsman radius, the flow is decelerated by
its own magnetic field and the process of accretion is switched to the diffusion regime.
Studies of the matter flow inside the Shvartsman radius presented by Bisnovatyi-Kogan
and Ruzmaikin
 \cite{Bisnovatyi-Kogan-Ruzmaikin-1974,
Bisnovatyi-Kogan-Ruzmaikin-1976}, have shown that in the region  $r < R_{\rm sh}$ the
spherical flow is transformed into the magnetic slab with small angular momentum. The
material of the slab is confined by the intrinsic magnetic field of the flow and is
moving towards the compact star as the field dissipation proceeds. Numerical simulations
of the magnetized accretion flow onto a black hole are presented in
\cite{Igumenshchev-etal-2003}.

A possibility for such a scenario to be realized in case of accretion onto a neutron
star has been recently discussed in   \cite{Ikhsanov-Beskrovnaya-2012,
Ikhsanov-Finger-2012}. In particular, formation of the magnetic slab has been shown to
occur under the condition $r_{\rm A} < R_{\rm sh}$, which is satisfied provided the
relative velocity of the neutron star with respect to the wind of its component is
$v_{\rm rel} < v_{\rm ma}$, where
    \be\label{vma}
 v_{\rm ma} \simeq 460\ \beta_0^{-1/5}\ \mu_{30}^{-6/35}\ \dmf_{15}^{3/35}\
 m^{12/35}\ \left(\frac{c_{\rm s}(r_{\rm G})}
 {10\,\text{km\,s$^{-1}$}}\right)^{2/5}\ \text{km\,s$^{-1}$}.
 \ee
The magnetospheric radius of the neutron star in this case ranges as
 $r_{\rm ma}
\leq r_{\rm m} \leq r_{\rm A}$, where
  \be\label{rma}
 r_{\rm ma} = \left(\frac{c\,m_{\rm p}^2}{16\,\sqrt{2}\,e\,k_{\rm B}}\right)^{2/13}
 \frac{\alpha_{\rm B}^{2/13} \mu^{6/13} (GM_{\rm ns})^{1/13}}{T_0^{2/13} \dmf^{4/13}}
 \ee
is a solution of the system of equations
 \be\label{syst1}
 \left\{
 \begin{array}{ll}
 \displaystyle\frac{\mathstrut \mu^2}{2 \pi r_{\rm m}^6} = \rho(r_{\rm m}) c_{\rm s}^2(r_{\rm m}) & \\
  & \\
 \dmf_{\rm in}(r_{\rm m}) = \displaystyle\frac{\mathstrut L_{\rm X} R_{\rm ns}}{GM_{\rm ns}} & \\
  & \\
  \dmf_{\rm in}(r_{\rm m}) = 2 \pi r_{\rm m} \delta_{\rm m} \rho_0 v_{\rm ff}(r_{\rm m}) & \\
  & \\
  \delta_{\rm m} = \left[\alpha_{\rm B} t_{\rm ff}(r_{\rm m}) D_{\rm B}(r_{\rm m})\right]^{1/2} & \\
  \end{array}
 \right.
 \ee
Here $\dmf_{\rm in}$ is the rate of plasma penetration from the slab into the neutron
star magnetic field,  $\delta_{\rm m}$ is the diffusion layer thickness at the
magnetospheric boundary,
 \be
D_{\rm B} = \frac{c k_{\rm B} T_{\rm i}(r_{\rm m})}{16 e B(r_{\rm
m})}
 \ee
is the Bohm diffusion coefficient and  $\alpha_{\rm B} < 1$ is a dimensionless
parameter. Finally, $k_{\rm B}$ is the Boltzmann constant and $T_{\rm i}(r_{\rm m})$ is
the ion temperature in the magnetopause plasma   \cite{Artymovich-Sagdeev-1979}.

The first equation of this system reflects that the magnetic pressure due to the stellar
field balances the accretion flow pressure at the magnetospheric boundary. The second
one is the continuity equation. It shows that in the process of stationary accretion,
the rate of plasma penetration from the slab into the magnetosphere is equal to the
accretion rate onto the stellar surface. The value of
 $\dmf_{\rm in}$ is to a large extent determined by an effective diffusion coefficient
 which, in its turn, depends on the stability of the magnetospheric boundary.
If it is  unstable with respect to interchange instabilities (such as Raleigh-Tailor
and/or Kelvin-Helmholtz), the rate of accretion flow penetration into the magnetosphere
at the radius $r_{\rm A}$ is comparable to the rate of mass transfer between the system
components \cite{Arons-Lea-1976, Elsner-Lamb-1977, Burnard-etal-1983}. If the
interchange instabilities of the magnetospheric boundary are suppressed (e.g. by the
magnetic field shear, see \cite{Ikhsanov-Pustilnik-1996}), the plasma entry into the
field is governed by anomalous (Bohm) diffusion.
The rate of plasma outflow from the slab to the magnetosphere at the radius
 $r_{\rm A}$ in this situation is substantially smaller than the mass transfer rate between the system components
 derived from the observed X-ray luminosity of a pulsar
\cite{Elsner-Lamb-1984}. The rate of plasma inflow to the magnetospheric boundary in
this case exceeds the rate of plasma penetration into the stellar magnetic field. This
results in accumulation of matter at the inner radius of the slab and increase of
external pressure on the stellar magnetic field. According to the first and the third
equations of system ~(\ref{syst1}), this leads to magnetospheric radius decreasing,
while the rate of plasma entry into the stellar magnetic field increases as   $\dmf_{\rm
in} \propto r_{\rm m}^{-13/4}$. The accretion process gets into a steady state as the
magnetospheric radius reaches  the value $r_{\rm ma}$, at which the rate of plasma
diffusion into the stellar magnetic field attains the rate of mass transfer between the
system components evaluated from the observed X-ray luminosity of a pulsar.

%%%%%%%%%%%%%%%%%%   Таблица 3 %%%%%%%%%%%%%%%%%%%%%%%%%%%

\begin{table}
\begin{tabular}{lccc}

 \noalign{\smallskip}
  \hline
\noalign{\smallskip}

 Name&\hspace{5mm}
$\dot{\nu}_{\rm sd}^{\rm (t)}/\dot{\nu}_{\rm sd}^{\rm obs}$ ~&~
$\dot{\nu}_{\rm sd}^{\rm (0)}/\dot{\nu}_{\rm sd}^{\rm obs}$ ~&~
$\dot{\nu}_{\rm sd}^{\rm (sl)}/\dot{\nu}_{\rm sd}^{\rm obs}$\\
 \noalign{\smallskip}
 \hline
 \noalign{\smallskip}
OAO~1657--415 & 0.02 & 0.06 & 2.1\\
Vela~X--1 &     0.003 & 0.08 & 2.7\\
4U~1907+09 &    0.005 & 0.24 & 7.2\\
4U~1538--52 &   0.001 & 0.04 & 1.3\\
GX~301--2 &     0.003 & 0.24 & 8.4\\
X~Persei &      0.005 & 0.09 & 2.7\\
\noalign{\smallskip}
  \hline
    \noalign{\bigskip}
  \multicolumn{4}{l}{$^*$~~$\dot{\nu}_{\rm sd}^{\rm (t)} = \dfrac{K_{\rm
sd}^{\rm (t)}}{2 \pi I}$; \hspace{5mm} $\dot{\nu}_{\rm sd}^{\rm (0)} =
\dfrac{K_{\rm sd}^{\rm (0)}}{2 \pi I}$; \hspace{5mm} $\dot{\nu}_{\rm sd}^{\rm
(sl)} = \dfrac{K_{\rm sd}^{\rm (sl)}}{2 \pi I}$}
 \end{tabular}
 \caption{The ratio of expected to observed spin-down rate\,$^*$}
 \bigskip
 \label{tab-3}
 \end{table}

%%%%%%%%%%%%%%%%%%%%%%%%%%%%%%%%%%%%%%%%%%%%%%%%%%%%%%%%%%%

The spin-down torque applied to the neutron star due to interaction between its dipole
field and the magnetic slab at the magnetospheric boundary is limited by the inequality
$K_{\rm sd} \leq K_{\rm sd}^{\rm (sl)}$, where \cite{Ikhsanov-etal-2013}
  \be\label{ksd3}
 K_{\rm sd}^{\rm (sl)} = k_{\rm t}\ \frac{\mu^2}{\left(r_{\rm ma} r_{\rm cor}\right)^{3/2}}
 \left(1 - \frac{\Omega_0}{\omega_{\rm s}}\right).
 \ee
The maximum spin-down rate of the neutron star expected within this scenario can be
evaluated, taking   $\Omega_0 \ll \omega_{\rm s}$ and solving inequality $|\dot{\nu}_{\rm
sd}| \leq |K_{\rm sd}^{\rm (sl)}|/(2 \pi I)$. As a result one finds $|\dot{\nu}_{\rm
sd}| \leq |\dot{\nu}_{\rm sd}^{\rm (sl)}|$, where
 \be
 |\dot{\nu}_{\rm sd}^{\rm (sl)}| = \frac{k_{\rm t} \mu^2}
 {2 \pi I \left(r_{\rm ma} r_{\rm cor}\right)^{3/2}}.
 \ee
The ratio of the maximum possible spin-down rate expected within different accretion
scenarios to the observed rate for a number of the best studied long-period X-ray
pulsars are presented in Table~\ref{tab-3}. The values are derived for the following
neutron star parameters: radius  $R_{\rm ns} = 10$\,km, moment of inertia $I =
10^{45}\,\text{g\,cm$^2$}$ and dipole magnetic moment $\mu = (1/2) B_{\rm CRSF} R_{\rm
ns}^3$, where $B_{\rm CRSF}$ is the magnetic field strength on the stellar surface
estimated through observations of the cyclotron line in the spectrum of a corresponding
pulsar. Normalization of parameter $\alpha_{\rm B}=0.25$ has been chosen by analogy with
solar wind penetration into the Earth magnetic field \cite{Gosling-etal-1991}. As
clearly seen from  Table~\ref{tab-3}, the spin-down rates expected for both a spherical
accretion, $\dot{\nu}_{\rm sd}^{\rm (t)} = K_{\rm sd}^{\rm (t)}/(2 \pi I)$ and
an accretion from a hot turbulent envelope, $\dot{\nu}_{\rm sd}^{(0)} = K_{\rm sd}^{(0)}/(2
\pi I)$, are significantly smaller than the observed values, while the spin-down rates
calculated for the case of accretion from a magnetic slab exceeds the observed values.
Observational results in this case can be fitted adopting the efficiency parameter value
in the range  $k_{\rm t} \sim 0.1 - 0.8$.

%%%%%%%%%%%%%%%%%%   Таблица 4 %%%%%%%%%%%%%%%%%%%%%%%%%%%

\begin{table}
\begin{tabular}{lccc}
 \noalign{\smallskip}
  \hline
\noalign{\smallskip}
 Name~ &~
 $r_0(\dot{\nu}_{sd}^{obs})$,\,cm~ &~
 $r_A(L_x)$,\,cm~ &~
 $r_0/r_A$ \\

 \noalign{\smallskip}
 \hline
 \noalign{\smallskip}

OAO~1657--415 & $1.4\times10^{8}$ & $6.8\times10^{8}$ & 0.20\\
Vela~X--1     & $1.3\times10^{8}$ & $5.6\times10^{8}$ & 0.24\\
4U~1907+09    & $2.9\times10^{8}$ & $6.0\times10^{8}$ & 0.48\\
4U~1538--52   & $9.5\times10^{7}$ & $6.3\times10^{8}$ & 0.15\\
GX~301--2     & $2.6\times10^{8}$ & $5.5\times10^{8}$ & 0.48\\
X~Persei      & $4.6\times10^{8}$ & $1.8\times10^{9}$ & 0.25\\
 \noalign{\smallskip}
  \hline
   \noalign{\smallskip}
 \end{tabular}
 \caption{Magnetospheric radii calculated using Eq.~(\ref{r0}) and observed spin-down rates of the pulsars}
 \bigskip
 \label{tab-4}
 \end{table}

%%%%%%%%%%%%%%%%%%%%%%%%%%%%%%%%%%%%%%%%%%%%%%%%%%%%%%%%%%%%%%%%%%%

 \section{Equilibrium period}\label{peq}

According to Eq.~(\ref{main}), a spin period of an accreting neutron star evolves
towards the equilibrium period,  $P_{\rm eq}$, defined through the balance of spin-up
and spin-down torques, exerted on a star from the accretion flow. The value of this
period depends on the parameters of both a binary system and its components as well as
on the accretion flow properties, i.e. its geometry and physical conditions in the
accreting material.

  \subsection{Accretion from a Keplerian disk}

The magnetosphere of a neutron star accreting matter from a Keplerian disk contains
both open and closed field lines. Angular velocity of the Keplerian disk material at the
radius $r < r_{\rm cor}$ exceeds the spin angular velocity of the star, $\omega_{\rm
s}$. Therefore, the interaction between the stellar magnetic field and the disk in this
spatial region leads to spin-up of the star. Interaction of the magnetic field with
outer parts of the disk located beyond the corotation radius ($r > r_{\rm cor}$) results
in deceleration of stellar rotation since the angular velocity of the matter in this
part of the Keplerian disk is smaller than the angular velocity of the star. Equilibrium
period of the star in this case can be evaluated from the equation   $|K_{\rm su}^{\rm
(Kd)}| = |K_{\rm sd}^{\rm (Kd)}|$, where \cite{Pringle-Rees-1972}
 \be
 |K_{\rm su}^{\rm (Kd)}| = \dmf \left(GM_{\rm ns} r_{\rm A}\right)^{1/2}
 \ee
is the spin-up torque, which is the product of specific angular momentum of the matter
located at the magnetospheric boundary,  $\sim r_{\rm A}^2 \omega_{\rm k}(r_{\rm A})$,
and the rate of accretion of this matter onto the stellar surface, whereas
\cite{Lynden-Bell-Pringle-1974}
 \be
 |K_{\rm sd}^{\rm (Kd)}| = \frac{\mu^2}{r_{\rm cor}^3}
 \ee
is the spin-down torque arisen due to interaction between the stellar field lines and
remote parts of the disk. Here  $\omega_{\rm k} = \left[r^3/(GM_{\rm ns})\right]^{1/2}$
is the Keplerian angular velocity. Solving this equation for  $P_{\rm s}$, one finds
\cite{Lipunov-1987, Alpar-2001}
 \be\label{peqkd}
P_{\rm eq}^{\rm (Kd)} \simeq 17\,\mu_{30}^{6/7}\,\dmf_{15}^{-3/7}\,m^{-5/7}\ \text{s}.
 \ee
Let us note that  the value of $P_{\rm eq}^{\rm (Kd)}$, calculated for the parameters of
long-period X-ray pulsars is significantly smaller than the observed spin periods of these
objects. This finding inevitably casts  doubt on the presence of a Keplerian disk in these
sources. Moreover, numerical simulations \cite{Urpin-etal-1998} have shown that expected
spin evolution of long-period X-ray pulsars in this case differs from observed. The
formation of a Keplerian accretion disk around a slowly rotating neutron star leads to
monotonous decrease of its spin period at a rather high rate. For example, the typical
spin-up time-scale of the neutron star
 GX\,301--2 within this scenario is only 10\,years
\cite{Ikhsanov-Finger-2012}. That is why the geometry of accretion flow in long-period
X-ray pulsars is usually considered in quasi-spherical approximation.

 \subsection{Quasi-spherical accretion}

Magnetosphere of the neutron star undergoing accretion from a quasi-spherical flow or a
hot turbulent envelope is closed and does not contain open field lines. Interaction
between the neutron star magnetic field and accretion flow in this case occurs in the
local region called magnetopause which is situated at the magnetospheric boundary of the
neutron star. Hence, the neutron star can spin down only provided
 $\omega_{\rm f}(r_{\rm A}) <
\omega_{\rm s}$, where
  \be\label{omegaf}
 \omega_{\rm f}(r_{\rm A}) = \xi \omega_{\rm f0} \left(\frac{r_{\rm G}}{r_{\rm A}}\right)^2
 \ee
is the angular velocity in the quasi-spherical flow at the magnetospheric boundary,
$\omega_{\rm f0} = \omega_{\rm f}(r_{\rm G})$ is the angular velocity of matter captured
by the star from the wind of its companion at the Bondi radius, and  $\xi$ is the
dimensionless parameter accounting for dissipation of angular momentum in the
quasi-spherical flow. Otherwise, as first noted by Bisnovatyi-Kogan
 \cite{Bisnovatyi-Kogan-1991}, interaction between the magnetosphere and material at its
 boundary will result in spin-up of the star. This means that evaluating the equilibrium
 period of the star in the case of quasi-spherical accretion one should account for not
 only the absolute value of torques applied to the neutron star but also for their sign.

Specific angular momentum of matter with angular velocity $\omega_{\rm f}(r_{\rm A})$ at the magnetospheric boundary
 is $r_{\rm A}^2 \omega_{\rm f}(r_{\rm A})$. The torque due
to infall of this matter from the magnetospheric boundary onto the stellar surface at the
rate  $\dmf$ can be evaluated \cite{Davidson-Ostriker-1973, Illarionov-Sunyaev-1975} as
 \be
 K_{\rm su}^{\rm (sp)} = \xi\,\dmf\,r_{\rm G}^2\,\omega_{\rm f0}
 \ee
and causes spin-up of the star.

The torque due to viscous tensions in the quasi-spherical flow at the magnetospheric
boundary in general case can be evaluated putting  $r_{\rm m} = r_{\rm A}$ into the
Eq.~(\ref{ksd2}) and taking into account that the turbulent velocity at the
magnetospheric boundary, $v_{\rm t}(r_{\rm A})$, in this expression is normalized on
its maximum value, a free-fall velocity,  $v_{\rm ff}(r_{\rm A})$. This yields
 \be
 K_{\rm visc} = \dmf\,\omega_{\rm s}\,r_{\rm A}^2
 \left(1 - \frac{\omega_{\rm f}(r_{\rm A})}{\omega_{\rm s}}\right)
 \left(\frac{v_{\rm t}(r_{\rm A})}{v_{\rm ff}(r_{\rm A})}\right).
 \ee
In case of accretion from a non-rotating ($\omega_{\rm f}(r_{\rm A}) = 0$) hot turbulent
 ($v_{\rm t} = v_{\rm ff}(r_{\rm A})$) envelope (scenario of a so called subsonic propeller
  \cite{Davies-Pringle-1981, Ikhsanov-2003,
Ikhsanov-2005}), this expression is reduced to the canonical value of the maximum
possible spin-down torque which is realized in the picture of quasi-spherical accretion,
$K_{\rm sd}^{(0)}$. In the scenario of spherical (Bondi) accretion  ($\omega_{\rm
f}(r_{\rm A}) = 0$, accretion from a free-falling flow), in which the turbulent
velocity does not exceed the linear velocity of matter in the magnetopause, $v_{\rm t}
\leq \omega_{\rm s} r_{\rm A}$, the value of spin-down torque, $K_{\rm visc}$, turns out
to be equal to $\mu^2/r_{\rm cor}^3$, which is consistent with the result, previously
obtained by Lipunov  \cite{Lipunov-1982}.

The equilibrium period of the star accreting material from a quasi-spherical flow can
be found solving equation  $K_{\rm su}^{\rm (sp)} = K_{\rm visc}$ for $P_{\rm s}$ and
normalizing the angular velocity of the matter captured by the neutron star at the Bondi
radius on the orbital angular velocity,  $\omega_{\rm f0} = 2 \pi/P_{\rm orb}$,
(here $P_{\rm orb}$ is the orbital period of the system) as
 \be
 P_{\rm eq}^{\rm (sp)} = \frac{P_{\rm orb}\,r_{\rm A}^2}{\xi\,r_{\rm G}^2}
 \left(\frac{v_{\rm t}(r_{\rm A})}{v_{\rm t}(r_{\rm A})+ v_{\rm ff}(r_{\rm A})}\right).
 \ee
The value of this period calculated under assumption  $v_{\rm t}(r_{\rm A}) = v_{\rm
ff}(r_{\rm A})$ for the parameters of the long-period pulsar GX\,301-2,
 \be\label{peqsp}
 P_{\rm eq}^{\rm (sp)} \simeq 13\,\text{s}\ \times\ \xi_{0.2}^{-1}\ B_{12}^{8/7}\ L_{37}^{-4/7}\ m^{-12/7}\ R_6^{20/7}
 \left(\frac{P_{\rm orb}}{41.5\,\text{d}}\right) \left(\frac{v_{\rm rel}}{400\,\text{cm\,s$^{-1}$}}\right)^4,
 \ee
is substantially smaller than the observed value (see Table~\ref{tab-1}). Here the
parameter  $\xi_{0.2} = \xi/0.2$ is normalized on its average value obtained in
numerical simulations of quasi-spherical accretion without accounting for the magnetic
field of the accretion flow \cite{Ruffert-1999}, $L_{37} = L_{\rm
X}/10^{37}\,\text{erg\,s$^{-1}$}$, $R_6$ is the neutron star radius in units of
$10^6$\,cm, and the velocity of the star in the frame of  surrounding gas is normalized on
the upper limit of the stellar wind velocity of its massive counterpart inferred from
observations \cite{Kaper-etal-2006}. Let us note that assumption about
a higher stellar wind velocity encounters problems while explaining high X-ray luminosity
of this source (expected mass capture rate by the neutron star under this assumption
turns out to be significantly smaller than the accretion rate onto its surface). As a
consequence, the modeling of long-period pulsars in terms of quasi-spherical accretion
is also rather complicated.

  \subsection{Accretion from a magnetic slab}

The torque applied to a neutron star undergoing accretion of matter onto its surface
from a magnetic slab can be presented (see above) as  the difference of two components.
The first of them is caused by accretion of matter with angular momentum onto the
stellar surface,
 \be
 K_{\rm su}^{\rm (sl)} = \dmf\,\Omega_0\,r_{\rm m}^2.
 \ee
and leads to spin-up of the star. The angular velocity of matter at the inner radius of
the slab, $\Omega_0$, can be approximated taking into account that magnetic field of the
accretion flow does not strongly influence the accretion process beyond the
Shvartsman radius. This implies that mass accretion in the region
 $R_{\rm sh} < r < r_{\rm G}$ occurs in the quasi-spherical regime. In this case the
 angular velocity of material at the Shvartsman radius can be estimated using
 Eq.~(\ref{omegaf}) as
 \be
 \omega_{\rm f}(R_{\rm sh}) = \xi\,\omega_{\rm f0} \left(\frac{r_{\rm G}}{R_{\rm sh}}\right)^2.
 \ee

The matter infall inside Shvartsman radius is fully controlled by the magnetic field of
the accretion flow and proceeds in the diffusion regime. This circumstance essentially
complicates the problem about the evolution of angular momentum in the magnetic slab,
i.e. in the region $r_{\rm m} < r < R_{\rm sh}$. A particular situation which is
analyzed in this paper is the case of solid-body rotation of the slab. In the frame of
this hypothesis the angular velocity of matter at the inner radius of the slab is
adopted equal to that at the Shvartsman radius,   $ \omega_{\rm f}(R_{\rm sh})$.

The torque component applied to the star by the magnetic slab, $K_{\rm sd}^{\rm (sl)}$,
describes interaction between the stellar magnetic field and the material located at the
magnetospheric boundary and is determined by Eq.~(\ref{ksd3}). The equilibrium period of
the neutron star in this case can be evaluated solving equation  $K_{\rm su}^{\rm (sl)}
= K_{\rm sd}^{\rm (sl)}$, which implies
 \be\label{peqsl}
 P_{\rm eq}^{\rm (sl)} = \frac{k_{\rm t}\,\mu^2\,c_{\rm s}^{8/3}(r_{\rm G})\,P_{\rm orb}}
 {\xi\,\beta_0^{4/3}\,v_{\rm rel}^{8/3} \left(\dmf\,r_{\rm ma}^{7/2}\,(GM_{\rm ns})^{1/2} + k_{\rm t}\,\mu^2\right)}.
 \ee
Solving this equation for $v_{\rm rel}$
 \be
 v_{\rm rel} = \frac{k_{\rm t}^{3/8} \mu^{3/4} c_{\rm s}(r_{\rm G})\,P_{\rm orb}^{3/8}}
 {\xi^{3/8} \beta_0^{1/2} P_{\rm s}^{3/8} \left(\dmf\,r_{\rm ma}^{7/2}\,(GM_{\rm ns})^{1/2} + k_{\rm t}\,\mu^2\right)^{3/8}},
 \ee
and adopting $\xi =0.2$, $c_{\rm s}(r_{\rm G}) = 10\,\text{km\,s$^{-1}$}$ and $\beta_0 =
1$, we find that the average  spin period measured in the long-period X-ray pulsar
GX\,301--2 can be fitted in the scenario of accretion from a magnetic slab provided the
neutron star velocity relative to the wind of its massive counterpart is $v_{\rm rel}
\sim 450\,k_{\rm t}^{3/8}\,\text{km\,s$^{-1}$}$, which is in good agreement with its
estimate inferred from observations  \cite{Kaper-etal-2006}.

 \section{Conclusions}\label{conclusions}

Results presented in this paper give evidence that the intrinsic magnetic field of the
matter captured by the neutron star from the wind of its massive companion can
essentially influence the accretion process in the long-period X-ray pulsars.
Traditional accretion scenarios (from a Keplerian disk or a quasi-spherical flow) can be
applied in this case only provided $R_{\rm sh} < r_{\rm A}$. Under this condition the
magnetic field of the accretion flow can be neglected. Otherwise ($r_{\rm A} < R_{\rm
sh}$), we come to a qualitatively new accretion picture. Formation of a non-Keplerian
magnetic slab expected in this case force us to reconsider the interaction of the
accretion flow with the stellar magnetic field. This leads to new evaluation of the
magnetospheric radius (Eq.~\ref{rma}), torques exerted on the neutron star by the
accretion flow  (Eq.~\ref{ksd3}) and equilibrium period of a pulsar (Eq.~\ref{peqsl}).
Numerical estimates of the spin-down rate of a neutron star and its equilibrium period
expected within this approach are consistent with those derived from observations
without additional assumption about significant deviation of the neutron star parameters
from canonical values.

It should be noted that difficulties arising in the modeling of X-ray pulsars in the
frame of Bondi accretion were mentioned by Arons and Lea   \cite{Arons-Lea-1976} at the
dawn of the age of the accretion theory. They have shown that magnetospheric radius of
the neutron star forming in the process of interaction between its dipole magnetic field
and free-falling material cannot be smaller than the canonical Alfven radius $r_{\rm
A}$. In this case the magnetosphere does not contain open field lines and forms two cusp
points located at the magnetic axis (see Fig.\,2 from \cite{Arons-Lea-1976} and Eq.~5
from \cite{Michel-1977}). Plasma entry into the magnetic field of the star in this
situation occurs in the direction perpendicular to the magnetic lines of force, and in
general case proceeds via diffusion. The rate of both the classical (Coulomb) and
anomalous (Bohm) diffusion of the spherical flow into the magnetic field evaluated by
Arons and Lea turned out to be a few orders of magnitude less than the accretion rate
onto the stellar surface estimated from X-ray luminosity of the pulsars   (see also
\cite{Elsner-Lamb-1984}). This result has given evidence that either there exists a
process which makes it possible for the accretion flow to enter the stellar magnetic
field at the rate, substantially exceeding the rate of diffusion, or the structure of
accretion flow in X-ray pulsars deviates from that realized in Bondi scenario. A rapid
entry of the accreting matter into the stellar field could occur if the magnetospheric
boundary were unstable with respect to interchange-type instabilities (in particular,
Raleigh-Tailor and/or Kelvin-Helmholtz instabilities). Such a situation is possible,
however, only in the high-luminosity pulsars  ($L_{\rm X} > 3 \times
10^{36}\,\text{erg\,s$^{-1}$}$ \cite{Arons-Lea-1976, Elsner-Lamb-1977}) with periods in
excess of   1000\,s \cite{Ikhsanov-Pustilnik-1996}. Otherwise, the interchange
instabilities of the magnetospheric boundary are suppressed.  The overwhelming majority
of  long-period X-ray pulsars known nowadays do not satisfy these criteria that makes
very difficult to describe these objects within the scenario suggested by Arons and Lea.

Similar obstacles are met when one considers  accretion from a Keplerian disk.
Relatively low efficiency of gas diffusion from the disk into the stellar magnetic field
and stability of the magnetospheric boundary were mentioned in \cite{Anzer-Boerner-1980,
Anzer-Boerner-1983, Malagoli-etal-1996}. The main problem of this approach is, however,
a relatively small value of the neutron star equilibrium period, $P_{\rm eq}^{\rm
(Kd)}$. Bearing in mind that in case of accretion from a Keplerian disk the time scale
of pulsar evolution to the equilibrium period constitutes only fractions of percent of
the X-ray source lifetime
 \cite{Urpin-etal-1998}, a probability to discover a long-period X-ray pulsar proves to be
 negligibly small. At the same time, the abundance of X-ray pulsars with  periods
 $P_{\rm s} \gg
P_{\rm eq}^{\rm (Kd)}$ significantly exceeds the number of short-period pulsars.

Account for the intrinsic magnetic field of the accretion flow in modeling of mass
transfer between the components of high-mass  binaries with wind-fed X-ray pulsars
greatly facilitate solving above mentioned problems. High rate of the flow entry into
the stellar magnetic field in this case is reached due to increase of neither the
effective area of interaction (as in case of interchange instabilities of the
magnetospheric boundary) nor the diffusion time (as in the model of Ghosh and Lamb
\cite{Ghosh-Lamb-1979}), but rather due to generation of large density gradient at the
magnetospheric boundary. The process of gas entry into the stellar magnetic field in
this case is caused by the same mechanisms as in the solar wind penetration into the
Earth magnetosphere (e.g. drift-dissipative instabilities of the magnetopause
\cite{Gosling-etal-1991, Artymovich-Sagdeev-1979}) and can be treated in terms of
anomalous (Bohm) diffusion. The expected spin evolution of pulsars undergoing accretion
from a magnetic slab is consistent with observed evolution of long-period X-ray pulsars.
This gives us compelling reasons to believe that further development of the approach
involving accretion from a magnetized stellar wind is very promising with regard to
constructing the model of energy release in the accretion-powered sources.

\begin{theacknowledgments}
This work was partly supported by Ministry of Science and Education (under contracts
8394 and 8417), RAS Presidium Program N\,21 ``Non-stationary phenomena in the
Universe'', RFBR under grant N\,13-02-00077 and President Support Program for Leading
Scientific Schools NSH-1625.2012.2.
\end{theacknowledgments}

\end{document}